\documentclass[12pt,onecolumn,draftclsnofoot]{IEEEtran}

\IEEEoverridecommandlockouts

\usepackage{algorithmic}
\usepackage[ruled,vlined]{algorithm2e}

\usepackage{amsthm,amssymb,graphicx,multirow,color,amsfonts}
\usepackage{epstopdf}
\usepackage[noadjust]{cite}
\usepackage{tikz}
\usetikzlibrary{arrows,calc}		
\usepackage{bbm} 
\usepackage{pdfpages}
\usepackage{tabulary}
\usepackage{multirow}
\usepackage{comment}
\usepackage{lipsum}
\usepackage{float}
\usepackage{subcaption}
\usepackage{rotating}

\usepackage{xcolor}  

\definecolor{myRed}{RGB}{255,0,0}  
\definecolor{myBlue}{RGB}{0,0,255} 

\usepackage{cite}
\usepackage{amsmath,amssymb,amsfonts}
\usepackage{algorithmic}
\usepackage{graphicx}
\usepackage{textcomp}
\usepackage{xcolor}
\def\BibTeX{{\rm B\kern-.05em{\sc i\kern-.025em b}\kern-.08em
	T\kern-.1667em\lower.7ex\hbox{E}\kern-.125emX}}

\usepackage[bookmarks=false]{hyperref}
\hypersetup{draft} 

\begin{document}
	\graphicspath{{./figures/}}

\title{Temporal Spectrum Analysis for \\ Multi-Constellation Space Domain Awareness\\
}
\author{\IEEEauthorblockN{Mansour Naslcheraghi, Gunes Karabulut-Kurt}\\
	\IEEEauthorblockA{\textit{Poly-Grames Research Center, Department of Electrical Engineering} \\
		\textit{Polytechnique Montréal, Montréal, Canada}}
}


\maketitle

\begin{abstract}
Space Domain Awareness (SDA) system has different major aspects including continuous and robust awareness from the network that is crucial for efficient control over all actors in space. The observability of the space assets, on the other hand, requires efficient analysis of when and how observed space objects can be controlled. This becomes crucial when real-world spatial dynamics are taken into account, as it introduces complexities into the system. The real-world dynamics can reveal the structure of the network, including isolated and dominant stations. We propose a Temporal Spectrum Analysis (TSA) scheme that takes into account a set of real-world parameters, including actual dynamics of the objects in space, to analyze the structure of a ground-space network that inherits temporal spectrum as the key element of design. We study the potential interactions between multiple constellations using TSA and conduct comprehensive real-world simulations to quantify the structure of the network. Numerical results show how the temporal spectrum of each satellite affects the intra- and inter-constellation network structure, including interactions between ground stations and constellations.
\end{abstract}


\section{Introduction}
As human activity in space has rapidly expanded, the space environment has become increasingly crowded and contested. With thousands of active satellites, countless pieces of debris, and new objects being launched regularly, the potential for collisions and accidents has grown significantly. Incidents such as the emergency collision avoidance execution by the European Aeolus satellite with Starlink-44 \cite{Aeolus_Starlink44_collision} and the rendezvous (only 58m) between OneWeb and Starlink \cite{OneWeb_Starlink_collision} have been mitigated urgently by OneWeb, which altered the orbit of its satellite to avoid incidents such as the Iridium-Cosmos collision \cite{Satellite_Collisions_Iridium}, which created at least 1000 pieces of debris larger than 10 cm, plus many smaller ones. On the other hand, the term "Space Domain Awareness" (SDA) emerged more recently as space became a contested domain involving not just physical objects but also geopolitical interests and potential threats \cite{SSA_to_SDA_Book}. SDA reflects a broader perspective, not just the tracking of space objects but also understanding the intentions and capabilities of different space actors, including constellations in operation. These capabilities are crucial for the sustainability of space, and they cover broad civil applications as well.

SDA mechanisms often rely on complex automated control systems to control satellites, spacecraft, and other space assets. To design such a network, it is important to understand the physical structure of the real-world ground-space network, which can help in designing more robust and resilient systems by facilitating precise planning before incidents, such as collision avoidance command executions or Telemetry, Tracking, and Control (TT \& C) operations. There have been a few attempts recently \cite{RealTime_Distributed_Tracking, ML_orbit_classification, temporal_frequency, DecisionMaking_Graph_SDA, Mission_Planning, SDA_Falco_Gunes} to address challenging problems in SDA. From exploiting vision-based sensors for real-time collision avoidance \cite{RealTime_Distributed_Tracking}, to predicting the full state of satellites by classifying orbital states in large LEO constellations \cite{ML_orbit_classification}, studying time series in the space-ground network based on correlated frequencies between ground stations \cite{temporal_frequency}, employing decentralized decision-making to handle connectivity and communication delays for efficient sensor tasking \cite{DecisionMaking_Graph_SDA}, and using Machine Learning (ML) tools to plan distributed, autonomous systems that adapt to unfamiliar situations \cite{Mission_Planning}. The key advantage, if possible, is always to have efficient state prediction, so better task planning can be realized \cite{SDA_Falco_Gunes}.

However, none of the existing works studied the actual time domain spectrum analysis, which is crucial to understanding the behavior of a multi-constellation system in real-time. The motivation behind this research is to explore how the structure of an arbitrary network can be inferred from the potential interactions between satellites and ground stations. How the system's spatial dynamics imply certain conditions that shape the influence of one constellation on another. These questions form the basis of this paper, where we introduce Temporal Spectrum Analysis (TSA) as a technique to address them. The TSA can reveal the temporal structure of ground-space networks and can be applied to designing efficient and long-term mechanisms for various purposes such as resource allocation, routing, network optimization, and decision-making.

\textit{Contributions}: We first introduce the temporal spectrum, which is a pulse function evolving over time due to the spatial dynamics of the object orbiting the Earth as the central body, and it has a semi-deterministic pattern that follows continuous pulses of a step function in certain intervals. This modeling captures the accurate visibility in the ground-space network where a given ground station visits certain satellites on a regular basis. To the best of our knowledge, such temporal spectrum modeling has no presence in the current state of the art. Given an arbitrary set of satellites along with their dedicated ground stations, we are interested in formulating the structure (wiring) of the network to study the correlations in spectrum patterns. For instance, we can identify ground stations that share simultaneous temporal spectra to the same satellites or have complementary coverage patterns. We apply eigenvalue decomposition to derive information on ground stations with strongly overlapping satellites' temporal spectra or isolated ones with less activity, which are major aspects of collective knowledge in a robust SDA system. We model possible interactions between multiple constellations, which can be conveyed via temporal spectrum patterns between constellations. This interaction is modeled by pulse intensity, which is a metric to quantify the potential strength of a given constellation in the observability of the target satellites, meaning that it will quantify the behavior of the network in observing satellites throughout time. The proposed framework is tested by a sophisticated in-house framework designed for real-time SDA simulations. The TSA technique is applied to a multi-constellation system composed of satellites from the constellations Starlink, OneWeb, and Iridium and provides insights into the network structure for each individual constellation and the multi-constellation system.

\section{System Model}
Let us start by defining the system of multiple constellations denoted by $C = \{ c_1, c_2, \dots, c_{|C|} \}$, where $|C|$ is the number of constellations, and constellation $c_i$ has its own dedicated ground stations denoted by the set $G^{c_i} = \{g_{1}, g_{2}, \dots, g_{|G^{c_i}|}\}$, where $|G^{c_i}| \le K$, $K \in \mathbbm{N}$ is the number of ground stations in constellation $c_i$ to handle a number of dedicated satellites denoted by the set $X^{c_i} = \{x_1, x_2, \dots, x_{|X^{c_i}|}\}$, where $|X^{c_i}| \le N$ for $N \in \mathbbm{N}$.
\subsection{Temporal Spectrum Modeling}
\label{subsec:groundStations}
\begin{figure}[!t]
	\centering
	\includegraphics[width=0.6\textwidth, trim=10 10 10 10, clip]{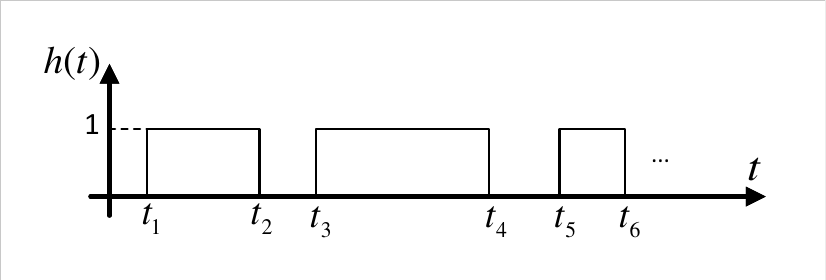}
	\caption{Temporal spectrum $h(t)$ vs. time $t$.}
	\label{fig:h(t)}
\end{figure}
\begin{figure}[!t]
	\centering
	\includegraphics[width=0.5\textwidth, trim=0 0 0 0, clip]{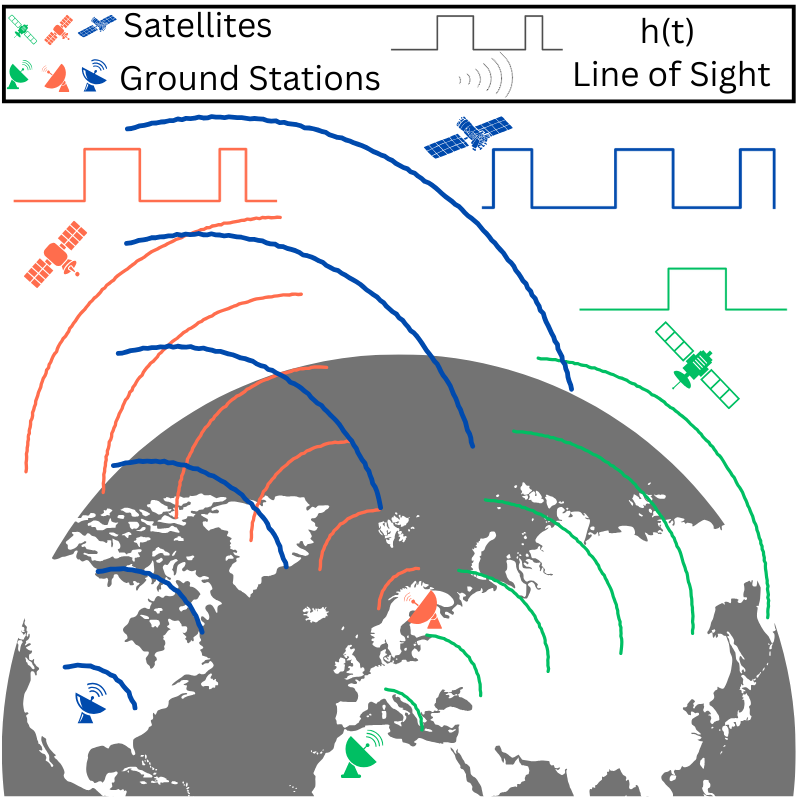}	
	\caption{Multi-Constellation system with temporal spectrum.}
	\label{fig:sys_model}
\end{figure}
Temporal spectrum is the actual physical time window from a ground station to a satellite on orbit, and for an arbitrary pair of satellite $x_{k} \in X^{c_j}$ and ground station $g_{i} \in G^{c_j}$, it follows a continuous pulse function denoted by $h_{ijk}(t)$, which has a shape similar to the one illustrated in Fig. \ref{fig:h(t)}. The pattern emerges from the spatial dynamics behavior of the system, and the control/data signals can be exchanged within this temporal spectrum. For an arbitrary satellite denoted by $x_k$, $t_m^{ijk}$ and $t_{m+1}^{ijk}$ are the start and end of one pulse, which we can also call the ``access window," and it has a length of $\Delta t_m^{ijk}$ in seconds. There is no overlap between pulses in $h_{ijk}(t)$ because $x_k$ never appears at two points in space at a particular time, meaning that the pulse exists only when $x_k$ is in line of sight with the ground station. Let us also explain why the pulse duration is different, as illustrated in Fig. \ref{fig:h(t)}. For instance, we can see that $t_2 - t_1 \ne t_4 - t_3 \ne t_6 - t_5$ for the same satellite w.r.t. the same ground station. This is due to the spatial dynamics of the system, including Earth’s rotation, satellite orbits, and gravitational fields. We have taken all these real-world features into account in our real-time simulation platform with a reliability of up to sixty days, and we will provide further details on this in the results section. Fig.~\ref{fig:sys_model} delineates an example of a multi-constellation system with temporal spectrum in action, where each satellite has a unique pattern of temporal spectrum to its dedicated ground station.

\section{Analysis}
\begin{figure}[!t]
	\centering
	\includegraphics[width=0.6\textwidth, trim=10 10 10 10, clip]{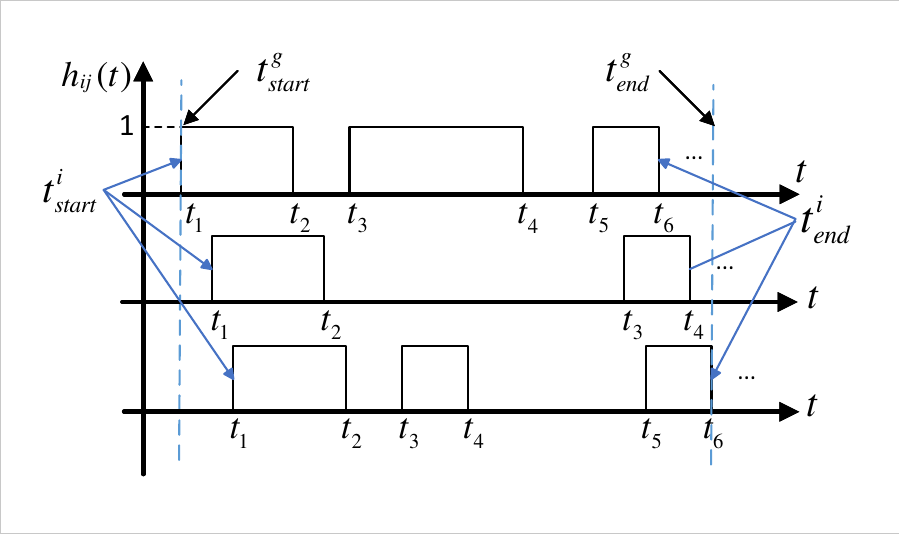}
	\caption{Simultaneity conditions between observations.}
	\label{fig:globalTime}
\end{figure}

A real-world ground-space network is dynamic, and the duration of line of sight for each satellite associated with its ground station varies between different satellites. Thus, to start analyzing the ground-space network, we need to first determine a start and end time in which all objects in the network have unanimous agreement on it, meaning that they are simultaneously operating within this time window. We call this window the ``simultaneity window,'' where every involved entity operates and forms a network in operation. Having this time window in hand, we can determine the size of the bitstream associated with each temporal spectrum, which ultimately facilitates the follow-up operations. To do this, let us first write the continuous-time temporal spectrum $h_{ijk}(t)$ as follows:
\begin{align}
	h_{ijk}(t) = & \sum_{m=1}^{N} \Big[ u\big(t - t_{m}\big) - u\big(t - t_{m} - \Delta t_{m}\big) \Big], \quad (a),
	\label{eq:h_ij(t)}
\end{align}
where statement $(a): t_{m+1} > t_m + \Delta t_{m}$ implies no overlap between pulses, \( t_{m} \) represents the starting time of the \( m \)-th pulse, \( \Delta t_{m} \) is the duration of the \( m \)-th pulse, \( N \) is the total number of pulses, and $t_{m+1} > \Delta t_{m}$ implies no overlap in the temporal spectrum. And \( u(t) \) is the unit step function defined as $u(t) =
\begin{cases}
	1, & t \geq 0, \\
	0, & t < 0.
\end{cases}$ Thus, the function $h_{ijk}(t)$ is a periodic sequence of pulses, each with an amplitude of 1, separated by certain gaps. It is worth noting that the indices in the time indicators are omitted for the sake of simplicity, since they are different as we described earlier. Let us define a unified global time \( T_g \), which spans from the earliest start time to the latest end time across all temporal spectram for all ground stations and satellites as $T_g = \left[ t_{\text{start}}^{\text{g}}, t_{\text{end}}^{\text{g}} \right]$, where 
\begin{equation}
t_{\text{start}}^{\text{g}} = \min\{ t_{\text{start}}^{i} \}, \quad t_{\text{end}}^{\text{g}} = \max\{ t_{\text{end}}^{i} \},
\label{eq:t_start_end}
\end{equation}
where $t_{\text{start}}^{i}$ and $t_{\text{end}}^{i}$ are start/end of first/last pulses, respectively. Fig. \ref{fig:globalTime} illustrates the graphical demonstration of the approach to determine the global time for three satellites visited by one ground station with varying pulse durations.
%
Now, let us form discrete-time of the function in eq. (\ref{eq:h_ij(t)}) w.r.t global time window as
\begin{equation}
	h_{ijk}[n] = 
	\begin{cases} 
		1 & \text{if }  n \in [t_m, t_m + \Delta t_m] \cap [t_\text{start}^i, t_\text{end}^i], \\
		0 & \text{otherwise}.
	\end{cases}
	\label{eq:h_ij[n]}
\end{equation}
In fact, eq. (\ref{eq:h_ij[n]}) is a binary sequence where each bit represents one second of information in the temporal spectrum, meaning that for a given time instance $t = t_0$, the binary value is 1 if the satellite $x_k$ is in line of sight with ground station $g_i$ in constellation $j$, and 0 otherwise. It is worth noting that one can tune the precision of sampling by $\alpha \ge 1$ seconds using $h_{ijk}[m] = h_{ijk}[\alpha \cdot m]$, subject to meeting Nyquist conditions, yet here we assume a non-sampled signal for the sake of simplicity.
Now, let us define a matrix $\mathbf{H}^{c_j}$ as the temporal spectrum strength between all ground stations in constellation $c_j$ and satellites $X^{c_j}$ as 
\begin{align}
	\mathbf{H}^{c_j} = \left[ \hat{h}_{i\,c_j\,x} \right]_{|G^{c_j}| \times |X^{c_j}|},
	\quad \text{for } i = 1,\dots,|G^{c_j}|,\notag \\ x = 1,\dots,|X^{c_j}|, 
	\label{eq:H_ij}
\end{align}
where each element $\hat{h}_{i{c_j}x}$ is the total time associated with temporal spectrum between ground station $g_i$ and satellite $x$ from constellation $c_j$ and it is specified by 
\begin{equation}
	\hat{h}_{i{c_j}x} = \sum_{n = t_\text{start}^g}^{t_\text{end}^g} h_{i{c_j}x}[n]  \quad \text{seconds},
	\label{eq:hat_h_ij}
\end{equation}
where $h_{i{c_j}x}[n]$ is given by eq. (\ref{eq:h_ij[n]}) and $ \hat{h}_{i{c_j}x} \le T_g$ since the overall observable window cannot exceed the global time.

\subsection{Intra-Constellation Dominant and Isolated Stations}  
Eigenvalue decomposition of the matrix \( \mathbf{H}^{c_j} \) reveals critical structural properties of the network. Large eigenvalues indicate redundant temporal spectrum patterns, while small eigenvalues highlight inefficiencies or limited \textit{access windows}. To analyze this, we examine \( \mathbf{J} = \mathbf{H}^{c_j} {\mathbf{H}^{c_j}}^T \), where \( {\mathbf{H}^{c_j}}^T \) is the transpose of \( \mathbf{H}^{c_j} \). This provides insights into satellite–ground station relationships and the overall robustness of the SDA network. In fact, the matrix \( \mathbf{J} \) quantifies the total \textit{access window} overlap or energy (strength) across ground stations in constellation \( c_j \). Diagonal entries \( \mathcal{J}_{ii} \) represent the observation strength of individual stations, where higher values indicate redundancy. Off-diagonal entries \( \mathcal{J}_{ij} \) (\( i \neq j \)) measure temporal spectrum overlap between stations \( g_i \) and \( g_j \), where large values illustrate redundant coverage patterns. Now, by decomposing \( \mathbf{J} \) through $\mathbf{J} = Q \Lambda Q^T$, where we obtain \( \Lambda \), a diagonal matrix of eigenvalues that corresponds to variance in observation modes, and \( Q \) is an orthogonal eigenvector matrix that captures interaction or redundancy among stations. The dominant and isolated ground stations \( g_{\hat{i}} \) are identified by
\begin{equation}  
	\hat{i} = \mathop{\arg \max}\limits_{i \in \{1, 2, \dots, |G^{c_j}|\}} \lambda_i, \quad 	\tilde{i}= \mathop{\arg \min}\limits_{i \in \{1, 2, \dots, |G^{c_j}|\}} \lambda_i. 
	\label{eq:hat{i}-tilde{i}}  
\end{equation}  
These stations, \( g_{\hat{i}} \) and \( g_{\tilde{i}} \), are pivotal for load balancing in constellation \( c_j \). The complete procedure is outlined in Algorithm \ref{alg:intra-constellation}.
\begin{algorithm}[!h]
	\caption{Dominant and Isolated Stations (Intra)}\label{alg:intra-constellation}
	\begin{algorithmic}[1] 
		\REQUIRE $G^{c_j}$, $X^{c_j}$, $t_m$, $\Delta t_m$, $t_\text{start}^i$, $t_\text{end}^i$
		\ENSURE dominant $\mathit{g}_{\tilde{i}}$ and isolated $\mathit{g}_{\hat{i}}$
		\STATE Solve (\ref{eq:t_start_end}) and get global time window $T_g$
		\STATE Initialize matrix $\mathbf{H}^{c_j}$
		\FOR{each station $g_i \in G^{c_j}$}
		\FOR{each satellite in $x_j \in X^{c_j}$}
		\STATE Get $\mathit{h}_{i{c_j}k}[n]$ using eq. (\ref{eq:h_ij[n]})
		\STATE Calculate $\mathit{\hat{h}}_{i{c_j}k}$ using eq. (\ref{eq:hat_h_ij}) and form $\mathbf{H}^{c_j}$  
		
		\ENDFOR
		\ENDFOR
		\STATE Calculate \( \mathbf{J} =  \mathbf{H}^{c_j} {\mathbf{H}^{c_j}}^T \)
		\STATE Solve $\det (\mathbf{J}-\lambda \mathbf{I}) =0$ and get $\lambda_i$
		\STATE Solve eqs. (\ref{eq:hat{i}-tilde{i}})
		\RETURN $\mathit{g}_{\tilde{i}}$ and $\mathit{g}_{\hat{i}}$
	\end{algorithmic}
\end{algorithm}
\subsection{Inter-Constellation Temporal Interaction}  
To quantify the role of each constellation in the temporal spectrum pattern, we introduce a metric that captures the interaction dynamics between constellations. Specifically, we define \textit{pulse intensity} as the frequency at which a typical satellite is accessed by a ground station. This metric embeds inter-constellation information flow and emergency response capabilities. The \textit{pulse density} \( \rho_{ij} \) measures how frequently satellites in constellation \( X^{c_j} \) are accessed by ground stations in constellation \( G^{c_i} \). It is defined as the number of pulses per unit time, irrespective of pulse duration:  
\begin{equation}  
	\rho_{ij} = \frac{ \mathcal{P}_{ij} }{T_g} \quad \text{pulses/second},  
	\label{eq:pulseDensity}  
\end{equation}  
where \( \mathcal{P}_{ij} \) is the total number of pulses given by
\begin{equation}
\mathcal{P}_{ij} = \sum_{n=t_{\text{start}}^i + 1}^{t_{\text{end}}^i} \left[ h_{i{c_j}j}[n] - h_{i{c_j}j}[n-1] \right]^{+},
\end{equation}
where \( [a]^{+} = \max(0, a) \) counts only positive transitions from 0 to 1. The distribution of pulses across constellations, namely the distribution of $\rho_{ij}$, follows a probability mass function (PMF) denoted by \( f_{i \rightarrow j}(k) \) that relies on the spatial dynamics of the system, where \( k = \left\lceil \frac{ \rho_{ij} }{60} \right\rceil \) is the number of pulses per hour for the interaction from constellation \( c_i \) to \( c_j \). $\left\lceil a \right\rceil$ gives the smallest integer greater than or equal to $a$. This PMF is derived empirically from real-time simulations, which capture the likelihood of observing \( \rho_{ij} \) pulses in a given time frame.
However, to quantify the structure of the network in terms of the pulses, we can identify the ground station \( g_i \) with the highest number of pulses for a given satellite and time interval. With this choice, we make the strongest selection for interaction from one constellation to another, which illustrates the strongest representation from a constellation, and it is defined by
\begin{equation}  
	P_{ij} = \max_{i \in \{1, \dots, |G^{c_i}|\}} \sum_{l=1}^{|X^{c_j}|} \mathcal{P}_{ij}.  
	\label{eq:pulseNumbers}  
\end{equation}  
Next, let us utilize \( P_{ij} \) and construct the \textit{interaction intensity} matrix \( \mathbf{P} \) as follows. 
\begin{align}
	\mathbf{P} = \left[P_{ij}\right]_{|C| \times |C|},
	\quad \text{for } i,j = 1, \dots, |C|,
	\label{eq:P}
\end{align}
where \( |C| \) is the number of constellations. To analyze the interactions between constellations, we again perform eigenvalue decomposition on \( \mathbf{P} \) as $\mathbf{P} \vec{v}_i = \gamma_i \vec{v}_i, \quad i = 1, 2, \dots , |C| $, where \( \gamma_i \) are the eigenvalues and \( \vec{v}_i \) are the corresponding eigenvectors. A large \( \gamma_{\max} \) indicates strong overall connectivity between constellations, while \( \vec{v}_{\max} \), associated with \( \gamma_{\max} \), reveals the relative contribution of each constellation to the dominant interaction mode.

\section{Numerical Results}
\begin{figure}[htbp]
	\centering
	\includegraphics[width=0.7\columnwidth]{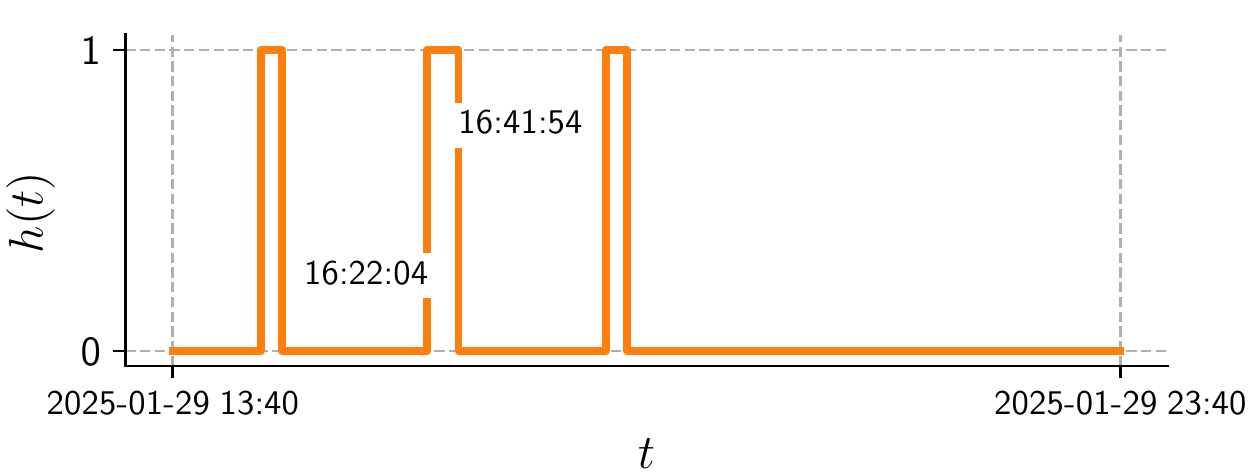}
	\caption{Temporal spectrum $h(t)$ for OneWeb-12 vs. ground station Alice Springs, Australia (ASA/ASN)..}
	\label{fig:oneweb_h(t)}
\end{figure}

\begin{figure}[!b]
	\centering
	\includegraphics[width=0.55 \textwidth]{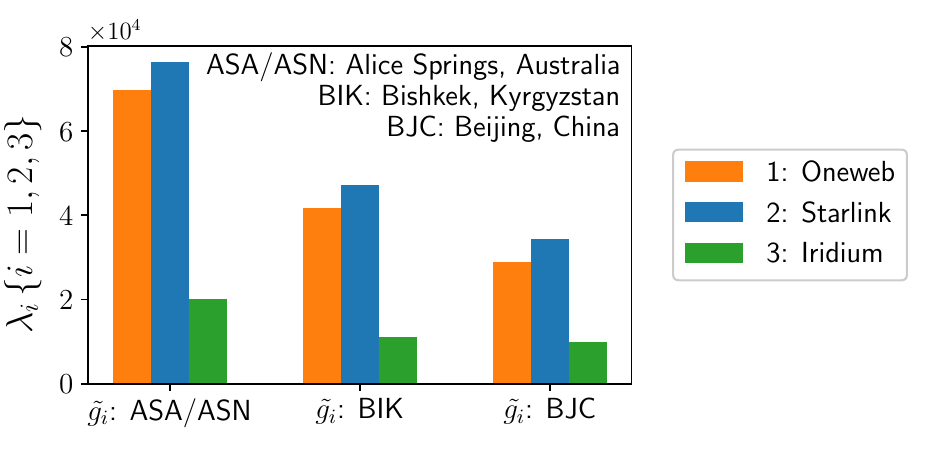}
	\caption{Top three dominant stations for all constellations.}
	\label{fig:dominant}
\end{figure}

\begin{figure}[!t]
	\centering
	\includegraphics[width=0.6 \textwidth]{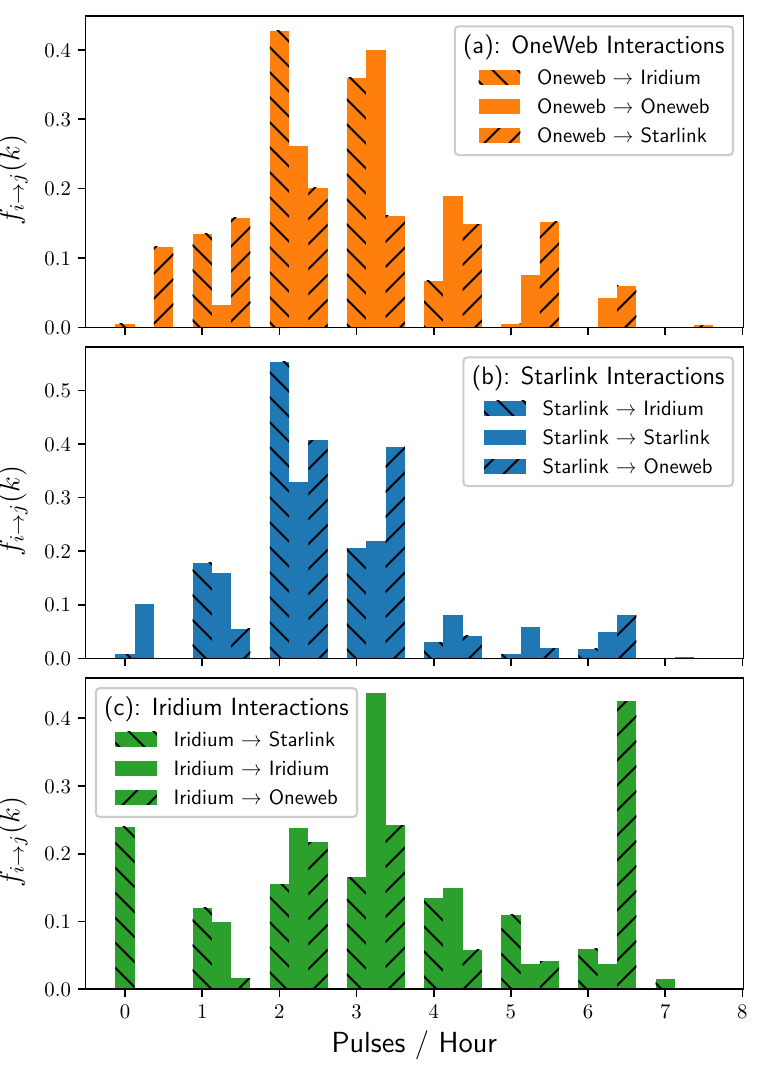}
	\caption{PMF of pulses per hour for different constellations.}
	\label{fig:oneweb_vs_others}
\end{figure}

Numerical results are conducted on an in-house real-time platform that integrates Simplified General Perturbations 4 (SGP4) \cite{SPG4_paper} and High Precision Orbital Propagator (HPOP) from the STK platform \cite{HPOP}. Satellites are sampled from the real-world data source CelesTrak \cite{CelesTrak} for the constellations OneWeb, Starlink, and Iridium. Full spatial dynamics are taken into account from the Two-Line Element (TLE) set \cite{CelesTrak}. First, let us illustrate the temporal spectrum in action, which is delineated in Fig. \ref{fig:oneweb_h(t)}. This example is for satellite OneWeb-12 versus the International Ground Station (IGS) Alice Springs, meaning that if OneWeb-12 were allowed to communicate with this IGS station, the demonstration shows the moments when visibility was possible. In practice, OneWeb is not using IGS, and this is just a sample scenario in our study, which is similar for every other pair of satellites and ground stations.
First, \textit{Scenario A}, to evaluate algorithm \ref{alg:intra-constellation}. We obtained dominant ground stations ($g_{\tilde{i}}$) and omit the demonstration of weak ground stations ($g_{\hat{i}}$) due to lack of space. We have deployed a network composed of the IGS network (39 stations), and the constellations OneWeb (651 satellites), Starlink (6909 satellites), and Iridium (30 satellites), with a precision of $\alpha = 1$ second and duration of 2 hours. Fig. \ref{fig:dominant} delineates the results where algorithm \ref{alg:intra-constellation} is executed for each constellation separately. It takes memory of size $\frac{|G^{c_j}| \times |X^{c_j}| \times (t^{g}_{\text{end}} - t^{g}_{\text{start}}) \times 3600}{\alpha}$ bits to build matrix $\mathbf{H}^{c_j}$. We demonstrate the top three stations which have the highest visibility to all constellations, yet with some differences in strength due to the nature of spatial dynamics between constellations.
\begin{table}[!t]
	\centering
	\begin{footnotesize}
		\caption{simulation parameters for scenario B}
			\begin{tabular}{c|c}
				\textbf{Parameters} & \textbf{	Values: Oneweb, Starlink, Iridium } \\ \hline
				Number of Satellites & 30, 50, 20 \\ 
				Ground Stations & 10, 20, 4 \\ 
				Altitude (Km) & $\sim$ 1200, 600 ,780 \\	
				Orbital Propagator & SGP4 and HPOP for all\\
				Period (minutes)    & $\sim$ 109, 90, 100\\
				Inclination (degree) & $\sim$ 89$^{\circ}$, 53$^{\circ}$, 86$^{\circ}$ \\
				Mean Motion (rev/day) & $\sim$ 13.15 ,15, 14.5 \\
				Simulation Duration (hours) & 10 
			\end{tabular}
		\label{tab:simParam}
	\end{footnotesize}
\end{table} 
Next, \textit{Scenario B}, to evaluate pulse distributions, namely PMFs $f_{i \rightarrow j}(k)$. Networks of ground stations in multiple countries across the globe are deployed to provide global coverage. Table \ref{tab:simParam} summarizes some of the key parameters for this scenario. Fig. \ref{fig:oneweb_vs_others} demonstrates the PMF of pulses for stations of different constellations against their own satellites and satellites from other constellations. We observe that for a given constellation, there is temporal spectrum available to another constellation with a variable number of pulses. We can also observe similarities for all constellations around 2–4 pulses on average with the highest probability. The reason is that for LEO orbits within 10 hours, satellites can make at least two complete orbits around the Earth. Let us focus on one particular behavior here. OneWeb and Starlink stations potentially have the highest interactions with Iridium satellites, as illustrated in Fig. \ref{fig:oneweb_vs_others} (a), (b), due to frequent lines of sight from these constellations to Iridium satellites. Similarly, Iridium has the highest probability of visiting satellites from OneWeb, as illustrated in (c). These behaviors show how we can infer the physical structure of a space-ground network and plan complicated tasks ahead of time.   
\begin{figure}[!t]
	\centering
	\includegraphics[width=0.3 \textwidth]{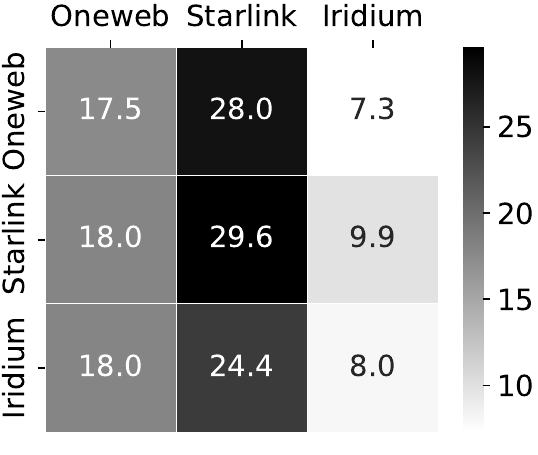}
	\caption{Interaction intensity matrix $\mathbf{P}$.}
	\label{fig:heatP}
\end{figure}
Finally, Fig. \ref{fig:heatP} demonstrates the interaction intensity between constellations, encoded by the \textit{interaction intensity} when the most frequent pulses are taken into account. This is the matrix $\mathbf{P}$ as in eq. (\ref{eq:P}), and we can see that Starlink has the highest pulse intensity, meaning that Starlink's ground stations visit its satellites more frequently. As for the inter-constellation interactions, Iridium has the lowest interaction values since it is a relatively small constellation with limited temporal spectrum available to other constellations. Obtaining eigenvalues and eigenvectors of the adjacency matrix \(\mathbf{P}\) reveals key insights into the visibility network of the three constellations. The dominant eigenvalue (\(\gamma_1 = 55.1\)) indicates strong overall connectivity, with the eigenvector showing that all three constellations contribute significantly, especially Starlink (\(0.62\)), which is the most influential one. The complex conjugate eigenvalues (\(\gamma_2 = -0.02 + 1.96j\), \(\gamma_3 = 0.02 - 1.95j\)) represent a cyclic visibility pattern with Iridium (\(0.74\)) playing a central role in these patterns, which persist over time due to their small real parts. Overall, the network exhibits robust connectivity with periodic visibility dynamics. One can also see Fig. \ref{fig:heatP} as the weights of a directed graph where each weight represents the interaction intensity from one constellation to another, which has a physical reality in the real world if constellations were to share this temporal spectrum with each other over their ground stations.
\section{Concluding Remarks}
Space Domain Awareness is crucial for sustainability, and due to increasing threats to space assets, space actors share responsibilities. In this paper, we have analyzed potential interactions not only within individual constellations but also between constellations. By analyzing the temporal spectrum, we demonstrated how we can derive information on the actual structure of the network along with potential interactions. These interactions are crucial for designing a robust SDA system that requires efficient handling of control signals. The temporal spectrum analysis reveals the temporal structure of the ground-space network and can be applied to designing efficient and long-term mechanisms for various purposes such as resource allocation, routing, network optimization, and decision making.


\bibliographystyle{IEEEtran}
\bibliography{ref}

\end{document}